 \documentclass[manuscript]{aastex}    

\shortauthors{Salo et al.}

\begin{document}


\title{Bars do drive spiral density waves}


\author{H. Salo\altaffilmark{1} and E. Laurikainen\altaffilmark{1,2}}
\altaffiltext{1}{Dept. of  Physics/Astronomy Division, University of Oulu, FI-90014, Finland}
\altaffiltext{2}{Finnish Centre for Astronomy with ESO (FINCA), University of Turku, V\"ais\"alantie 20, FI-21500 Piikki\"o, Finland}
\author{R. Buta\altaffilmark{3}}
\altaffiltext{3}{Dept. of Physics and Astronomy, University of Alabama, Box 870324, Tuscaloosa, AL 35487}
\and
\author{J. H. Knapen\altaffilmark{4,5}}
\altaffiltext{4}{Instituto de Astrof\'\i sica de Canarias, E-38200 La Laguna, Tenerife, Spain}
\altaffiltext{5}{Departamento de Astrof\'isica, Universidad de La Laguna, E-38205 La Laguna, Tenerife, Spain}




\begin{abstract}

Recently, Buta et al. (2009, AJ, 137, 4487) examined the question "Do
Bars Drive Spiral Density Waves?", an idea supported by theoretical
studies and also from a preliminary observational analysis
\citep{block2004}. They estimated maximum bar strengths $Q_b$, maximum
spiral strengths $Q_s$, and maximum $m$=2 arm contrasts $A_{2s}$ for
23 galaxies with deep AAT $K_s$-band images. These were combined with
previously published $Q_b$ and $Q_s$ values for 147 galaxies from the
OSUBSGS sample and with the 12 galaxies from \cite{block2004}.  Weak
correlation between $Q_b$ and $Q_s$ was confirmed for the combined
sample, whereas the AAT subset alone showed no significant
correlations between $Q_b$ and $Q_s$, nor between $Q_b$ and
$A_{2s}$. A similar negative result was obtained in \cite{durbala2009}
for 46 galaxies.  Based on these studies, the answer to the above
question remains uncertain.

Here we use a novel approach, and show that although the correlation
between the {\it maximum} bar and spiral parameters is weak, these
parameters do correlate when compared {\it locally}.  For the OSUBSGS
sample a statistically significant correlation is found between the
local spiral amplitude, and the forcing due to the bar's potential at
the same distance, out to $\approx$1.6 bar radii (the typical bar
perturbation is then of the order of a few percent).  Also for the
sample of 23 AAT galaxies of \cite{buta2009} we find a significant
correlation between local parameters out to $\approx$1.4 bar radii.
Our new results confirm that, at least in a statistical sense, bars do
indeed drive spiral density waves.


\end{abstract}






\section{Introduction}

The response of an outer disk to a rotating bar is intimately related
to the maintenance of long lasting spiral arms in galaxies.  Optical
photometry \citep{schweizer1976} established that besides the gas and young
stars, the spirals are present also in the old population. This was
confirmed by near-infrared surveys \citep{esk2002},
and is particularly true for grand-design spirals \citep{knapen1996}.  Short-lived stellar density waves can be induced via disk
instabilities \citep{bertin1977,tremaine1978}, or by galaxy interactions
\citep{toomre1972}, but such transient patterns fade after $\sim$10
galaxy rotations \citep{sellwood1984}, unless they are maintained by
some feedback cycle, e.g. due to the swing amplification
\citep{toomre1981}. On the other hand, spiral arms are excited by a
growing bar, as demonstrated by the very first N-body simulations
\citep{hohl1971} and by analytical calculations \citep{atha1980}.  


Support for the bar/spiral connection is provided by the examples
where prominent spirals extend from the ends of the bar (see e.g. NGC
1300, p. 525 in Binney \& Tremaine 2008, or NGC 986 in
Buta et al. 2010).  Also, grand-design spirals are more frequent in
barred than in non-barred galaxies \citep{elm1982}.
Nevertheless, a direct connection between bars
and spirals has been difficult to prove observationally.

The possible driving of spirals by bars was addressed by
\citet{block2004}, who compared the maximum of bar-related torque
strength $Q_b$ (the maximum of tangential force amplitude normalized
by mean radial force) with the maximum associated with the spirals
,$Q_s$, after separating the bar and spiral components based on their
Fourier density amplitude profiles. Near-IR observations were used, 
to show the effect of the bar on the surrounding mass, rather
than on the gas or the formation of young stars.  Based on 12 galaxies
observed in the $K_s$-band, \citet{block2004} found a strikingly clear
correlation between $Q_b$ and $Q_s$. However, the correlation is less
clear in later studies using larger samples and deeper images. In
\cite{buta2005} we analyzed the $H$-band images from the Ohio State
University Bright Spiral Galaxy Survey (OSUBSGS, Eskridge et al. 2002)
for 147 galaxies, and were able to ``weakly verify a possible
correlation between $Q_s$ and $Q_b$''. More recently, in
\citet{buta2009} we analyzed deep $K_s$ Anglo-Australian Telescope
(AAT) observations for 23 galaxies, with no statistically significant
correlation between $Q_b$ and $Q_s$, nor between $Q_b$ and $A_{2s}$
(the maximum of m=2 Fourier density amplitude of spirals).  Similarly,
\cite{durbala2009} found no correlations when analyzing Sloan $i$-band
data for 46 isolated barred galaxies; a lack of correlation
between bar and spiral arm strengths was seen also by \cite{seiger2003} who
analysed 41 galaxies.  Nevertheless, \citet{buta2009} showed that the
correlation is present when combining the AAT data with the previous
data sets of \citet{block2004} and \citet{buta2005}.

In this paper the bar/spiral connection is re-investigated using the
same samples which were used by \cite{buta2005,buta2009}: the OSUBSGS
sample containing $\sim$100 barred galaxies, and our AAT sample of 23
barred galaxies. A novel approach is used: instead of comparing the
{\em maximum} bar strength with the {\em maximum} spiral density
amplitude, we compare the locally measured bar forcing and spiral
amplitude as a function of distance.  
The locally felt forcing due to bar is
a more important parameter than the maximum forcing, since
the $Q_b$ is typically attained well inside the spiral structure. 
Also, we examine the spiral density rather than force amplitude,
since the former measures more directly the possible {\em response} to
bar forcing.  Using our approach, a statistically significant
correlation is demonstrated to exist between the bar forcing and the
spiral amplitude, up to a considerable distance beyond the end of the
bar.

\section{Calculation of bar forcing and spiral amplitudes}

We calculate {\em the amplitude of the bar tangential forcing} as a
function of distance from the galaxy center, $Q_{bar}(r)$, and
compare it to the $m$=2 {\em surface brightness amplitude of the
  spirals}, $A_2(r)$, at the same radial distances, normalized to bar
radius, $r/R_{bar}$.  The Spearman rank correlation coefficient $r_s$
between $Q_{bar}$ and $A_2$ and its significance level are then
measured as a function of $r/R_{bar}$.  A non-parametric test is used
in order to avoid making assumptions about the distributions of the
compared quantities.  The distance up to which a significant
correlation is found is regarded as a statistical estimate of the
region inside which bars are able to drive spiral structure.

We use near-IR images to evaluate the galaxy potential, and derive the
tangential force amplitude at each distance, normalized to the
azimuthally averaged radial force 

$$Q_{bar}(r) = \frac{\max(|F_T(r,\varphi)|)}{<|F_R(r,\varphi)|>}.$$

\noindent
Several assumptions are made:
(1) the mass-to-luminosity ratio ($M/L$) is constant,
(2) the vertical profile of the disk and the bar is approximated with
an exponential function, and (3) the vertical scale height, $h_z$,
scales with the galaxy size as $h_z=0.1R_{K20}$ \citep{spelt2008}, where $R_{K20}$ is
the $m_K=20$ isophotal radius from 2MASS \citep{skr2006}. 
The calculations are made with polar integration
 \citep{salo1999,laurikainen2002}, based on azimuthal Fourier
 decomposition of the deprojected image

$$I(r,\varphi) = A_0(r) \left[1+\sum_{m=1}^{\infty}A_m(r)\cos\left(m\left[\varphi-\varphi_m(r)\right]\right)\right],$$

 \noindent which also provides the $m=2$ Fourier density profile used
 to characterize the spiral amplitudes (note the normalization). The
 gravity is calculated separately for each $m$ component (using
 $m=0,2,4,6,8,10$) and added together.  Compared to direct Cartesian
 integration, the polar method suppresses effectively the spurious
 force maxima which otherwise could arise in the noisy outer disks
 \citep{salo2004}.  To account for the different 3D-distribution
 of the bulge light, multi-component decompositions are used
 \citep{laurikainen2005}, reducing the artificial tangential force
 amplitudes arising from the bulge deprojection stretch.  The polar
 method makes it also easy to exclude the contribution of spirals on
 the calculation of $Q_{bar}$, by setting the $m>0$
 Fourier density amplitudes to zero beyond a certain cutting distance,
 $R_{cut}$, representing the end of the bar.

Our method is illustrated in Figure 1, where the radial profiles of
the $m$=2 and $m$=4 Fourier density amplitudes, and the
$Q_{bar}$ ratio are shown for NGC 1566.  In this particular
example the $A_2$ and $A_4$ have two well-separated local
maxima associated with the bar and spiral, but the distinction is not
always clear.  The $Q_{bar}$ profile is constructed both for the total
(bar+spiral) force field, and without the contribution of the spiral
arms.  The bar-only profile is obtained using a cutting distance
$R_{cut}=R_{bar}$ in the force calculation.  Our method of isolating
the bar forcing is different from that of \citet{buta2003,buta2005,
buta2009}, who extrapolated the bar density into the spiral region based
on Gaussian fits to Fourier density profiles. Here we assume that the
bar and the spiral dominate their radial domains with no significant
overlap.

\section{Bar driven spiral structure in the OSUBSGS and AAT samples}


The OSUBSGS is a magnitude-limited sample ($m_B<12.0$
mag) of galaxies with Hubble types $0\le T\le9$.  The $H$-band images
typically reach $20$ mag/arcsec$^2$ in depth.  Our
bar identifications and lengths are from Table 3 in
\cite{laurikainen2004a}, and are based on the Fourier amplitude and
phase profiles ('Fourier bars'; NGC 2207 was omitted, leaving 103
galaxies for analysis).

A statistically significant correlation is found between the amplitude
$A_2$, and the bar forcing, $Q_{bar}$, when these parameters are
compared at the same radial distances.  See Fig. 2 where the different
panels represent examples of measurements at successively larger
radial distances with respect to the bar. The bar force is cut using
$R_{cut}=R_{bar}$, which means that we are conservative in eliminating
any contamination by the spiral arms themselves in the forcing.  The
correlation is very strong just beyond the bar, and stays
statistically significant until $\sim$1.6$R_{bar}$ (rank correlation
coefficient $r_s=0.25$, significance $p=0.008$).  The correlation
is similar for long ($R_{bar}/h_R>1$) and short ($R_{bar}/h_R<1$)
bars, when normalized to the disk scale length.  The range of the
significant correlation is similar also for early ($T \le3$) and late
type ($T \ge4$) spirals (not shown in the plots).

Figure 3 collects the correlation coefficients between $A_2$ and
$Q_{bar}$ at different distances $r/R_{bar}$.  Note that in the bar
region, where $A_2$ represents the bar itself, the correlation is
strong as expected.  Outside the bar $A_2$ arises from spiral
structures (the forcing is still dominated by the bar).  The radial
trend depends only slightly on the adopted cutting distance for the
bar (using reasonable values $R_{cut}/R_{bar}=1.0-1.2$). 
The correlation is also insensitive to the exact assumptions made
about vertical structure: an uncertainty by a factor of two in $h_z$
corresponds to $\sim20\%$ uncertainty in bar strength. Monte Carlo
trials show that even $20\%$ random errors in $Q_{bar}$ (and $A_2$)
affect only marginally the significance of the correlation (for $10,
000$ trials the median $p=0.023$ for $r=1.6R_{bar}$).


We made a similar analysis for the AAT sample of 23 barred galaxies,
using the data and bar lengths from \cite{buta2009}.  The $K_s$-band
images typically reach a surface brightness level of 22-24
mag/arcsec$^{2}$. The sample covers a wide range of bar strengths,
extending to strongly barred galaxies, for which case a strongest correlation
between $Q_{bar}$ and $A_2$ is expected.  However, as
discussed in \cite{buta2009}, no statistically significant correlation
is obtained between the maximum of $Q_{bar}$ and the maximum of $A_2$
(Fig. 4a). The result is similar for galaxies having the maximum
spiral amplitude nearer ($R_{2s}<1.5 R_{bar}$) or further
($R_{2s}>1.5R_{bar}$) outside the bar.  However, a statistically
significant correlation is found if the {\em local} bar forcing at the
location of the maximum spiral amplitude is examined (Fig. 4b). The
correlation is particularly clear if we limit to cases where
$R_{2s}<1.5R_{bar}$. This is in accordance with our result 
for the OSUBSGS where a statistical dependence is present but
discernible only up to certain distance beyond the bar end. Indeed, if
we repeat the analysis we made for the OSUBSGS sample (Fig. 4c), a
statistically significant correlation is found between $Q_{bar}(r)$
and $A_2(r)$ up to 1.4 $R_{bar}$.  The somewhat smaller range in the
AAT analysis is probably due to the smaller sample size in comparison
to the OSUBSGS sample.

\section{Discussion}

\subsection{Comparison to previous studies of bar/spiral correlation}

Our analysis for OSUBSGS has indicated a significant correlation
between the {\em local} tangential bar force $Q_{bar}(r)$ and the {\em
  local} surface brightness Fourier amplitude $A_2(r)$, up to
1.6$R_{bar}$.  How does this compare with the previous analyses of
$Q_b$ vs.  $Q_s$, some of which indicated a correlation, while others
did not?  In \cite{buta2005} we did not explicitly state the
correlation coefficient, but using the data tabulated in the paper one
finds a Pearson linear correlation coefficient $r=0.35$. For a sample
of $N=146$ galaxies this implies a very significant correlation
($p=0.8\cdot10^{-5}$).  Though different quantities are compared,
this statistically significant correlation between maximum values
agrees with our present analysis of the correlation between local
quantities.

Likewise, the negative results in \cite{buta2009} for the AAT sample
alone (N=23), and in \cite{durbala2009} for the Sloan sample (N=46),
are accounted for by the smaller number of galaxies. The standard
procedure for testing the significance of a positive linear
correlation (see e.g. Wall \& Jenkins 2003, Sect.~4.2.2) implies that
in order to accept the correlation (at a significance level $p<0.01$),
the sample correlation coefficients must be $r>0.48$ and $r>0.34$, for
$N=23$ and $N=46$, respectively. Assume for a while that a linear
correlation between $Q_b$ and $Q_s$ exists, and that these quantities
are drawn from a bivariate-Gaussian distribution with an actual
correlation coefficient $\rho=0.35$ (equal to the sample correlation
coefficient in Buta et al. 2005).  We may then ask what the odds are
of detecting this correlation, i.e. of observing a sufficiently high
sample correlation when drawing a random sample with different
N's. Applying the Fisher probability distribution for the sample $r$
with a known $\rho$ \citep{wall2003} implies that for $N=23$ there is
only a 25\% chance of detecting the correlation, even for $N=46$ the
change is only about $50\%$. If we use the rank correlation
coefficient instead of the linear correlation coefficient, the chances
are reduced to 15\% and 38\%, respectively (obtained by Monte Carlo
trial estimates). Thus the negative results for these small samples do
not rule out true correlations.

The current method seems to be capable of exposing the correlation
even for fairly small samples, indicating the advantage of comparing
the local quantities.  In particular, close to the bar the spiral
amplitudes are strongly correlated with the bar forcing. This radial
dependence of the correlation probably explains the very strong
correlation found in \cite{block2004}.  The fact that their small
sample (N=12) showed the strongest correlation (r=0.86) is likely to
result from the way the galaxies were selected, containing many
examples where the spirals are strong right at the ends of the bar.
Indeed, according to the tabulated values in their paper, the mean
ratio between the radial locations of the spiral and bar maximum
forces was $<R_s/R_b>=2.5$, compared to $<R_s/R_b>=4.6$ in
\cite{buta2005}, which was based on a magnitude
limited sample of spirals, with many different types of
bar/spirals.

\subsection{Physical mechanisms}

Our observational analysis indicates a clear statistical
relation between bars and spirals. At least the following mechanisms
might account for this:

1) Spiral structure is a direct response to bar forcing and/or the
spirals represent a continuation of the density wave associated with
the bar \citep{lindblad1960,toomre1969, bertin1996}.  Recently, spiral
arms have also been interpreted as manifold orbits emanating from
unstable Lagrangian points near the bar ends
\citep{romero2006,atha2009}.  

2) Spirals are coupled to the bar via non-linear resonance coupling
\citep{tagger1987,masset1997}. Such couplings are
seen in N-body simulations \citep{rauti1999}, but it is uncertain
how frequent such cases really are.

The first explanation is likely to apply to the strong correlation
just outside the bar. In this case the spirals are expected to
share the constant (or slowly-evolving) pattern speed of the
bar. Similarly, in the case of non-linear coupling, though the spiral
pattern speed is slower than that of the bar, it still represents a
steady long-lived pattern.  On the other hand, in the region where no
correlation is present, the spirals
are independent structures, representing either a
long-lasting mode with a slower pattern speed than the bar
\citep{sellwood1988}, or are just short-lived transient wave packets with
a range of propagation speeds \citep{sellwood1991,salo2000}.

Note that some correlation may be present even if bars and spirals are
independent, since both types of structures are favored by
gravitationally more reactive disks. Nevertheless in this case a
correlation between maximum values would also be expected.

\cite{sellwood2008} discusses the importance of distinguishing between
long-lived spiral modes and transient waves. This stems from the fact
that the latter are much more efficient in driving angular momentum
transport in the disk. In case of steady patterns the angular
momentum exchange with stars is limited to resonances, whereas in the
case of transient patterns this may occur over a large range of radii.
Multiple transient patterns also lead to radial mixing of stars,
as well as secular heating of the disk.

In the current samples no significant correlation is seen beyond
$\sim1.6R_{bar}$, at which distance the typical force amplitudes
associated with the bar fall below a few percent level. This 
provides an observational lower estimate for the extent of
bar-driving. Namely, in the outer disk the determination of the spiral
amplitude is more prone to uncertainties due to image noise and
background subtraction, diluting any correlation that may be present.
Also, we are fairly conservative in cutting the bar contribution
close to $R_{bar}$, since in the case where the spiral arms are a part
of the mode associated with the bar, the local spiral forcing would
also contribute to maintaining the pattern.

Although the correlation we find is strictly statistical, it is
interesting to consider how it applies to the individual example of
NGC 1566, shown in Fig. 1. For this galaxy $R_{bar}=40\arcsec$, and
the $A_2$ maximum related to the strong spirals is at
$R_{2s}=68\arcsec$ \citep{buta2009}, though there are additional
spiral arms even beyond $100\arcsec$.  The strong spirals, with a
maximum at 1.7 bar lengths, fall marginally in the region where
bar-driving is expected.  Continuing the same interpretation, the
spiral beyond $\sim100\arcsec$ would be an independent
pattern.  Indeed, NGC 1566 has been quoted as an example of a galaxy
with at least two separate spiral structures \citep{bosma1992}.


\section{Conclusions}

A connection between bar forcing and spiral density amplitudes was
investigated for two near-infrared galaxy samples: 103 barred galaxies
from the magnitude-limited OSUBSGS survey, and
23 barred galaxies in our AAT Survey. 
The main results are:

(1) {\em For both samples a statistically significant correlation is found
between the local tangential bar forcing, $Q_{bar}(r)$, and the local spiral
amplitude, $A_2(r)$, up to a radial distance of $r\sim$1.5$R_{bar}$. }

\noindent The correlation suggests that, at least in a statistical
sense, the stellar spirals of the disk are not transient features but
rather represent a continuation of the bar mode itself, or are driven
by the bar through some mechanism.  Further out, the spirals may be
either independent modes or transient wave packets.

(2) {\em The obtained range of the correlation is
  similar for early and late-type spirals ($0\le T\le3$
  and $4\le T\le9$) , and also for small and large bars
  ($R_{bar}/h_r$ smaller/larger than unity).}

\noindent This does not favor the idea that only certain types of bars
could drive spiral structure, or that the forcing on the stellar
component requires the presence of significant gas
component. Nevertheless, the current samples are small, and a larger
number of galaxies is needed to draw reliable conclusions about
morphological type dependencies, or to probe whether the statistical
correlation extends to even larger distances beyond the bar.  In this
respect the forthcoming S$^4$G survey (The Spitzer Survey of Stellar
Structure in Galaxies; Sheth 2009, Sheth et al. 2010) will be
extremely useful, providing unprecedentedly deep 3.6 and 4.5 micron
observations for nearly 2300 nearby galaxies.



\clearpage

\begin{figure} 
\includegraphics[angle=0,scale=.75]{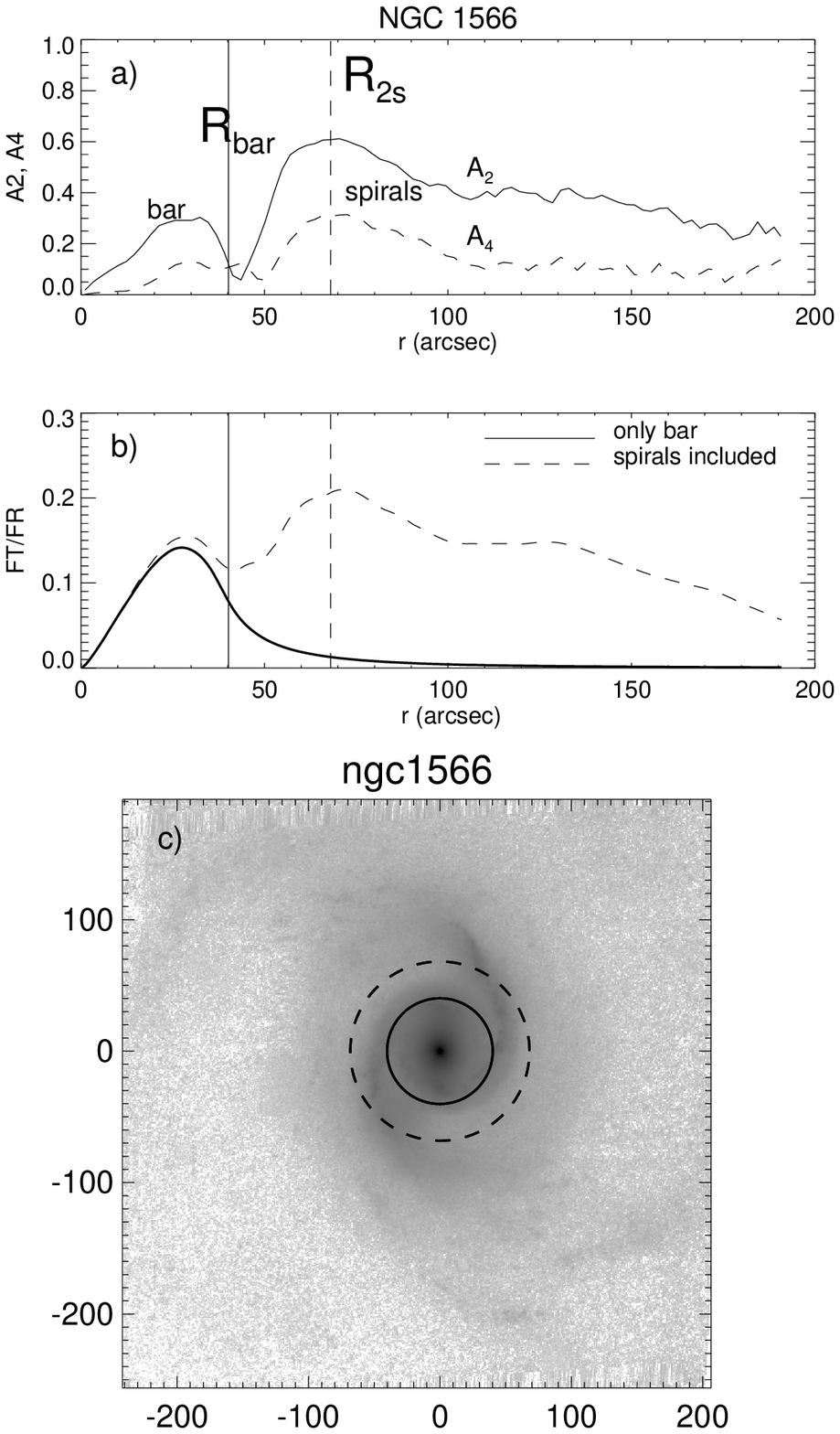}
\caption{ Barred galaxy NGC 1566, with strong outer spirals. In a) the
  normalized $A_2$ and $A_4$ Fourier amplitudes are shown as a
  function of radius. The two local maxima are associated with the bar
  and the spiral: the vertical lines indicate the bar length
  $R_{bar}$, and the distance of maximum $A_2$ of the spirals
  ($R_{2s}$). b) The calculated tangential forcing $Q_{bar}=FT/FR$,
  with and without the contribution from spiral arms: in the latter
  case the $m>0$ Fourier components have been set to zero for
  $r>R_{cut}$, here using $R_{cut}=R_{bar}$. c) The deprojected
  $K_s$-band image \citep{buta2009}, together with circles of radii
  $R_{bar}$ and $R_{2s}$.
\label{fig1}
}
\end{figure}

\clearpage
\begin{figure} 
\includegraphics[angle=0,scale=.90]{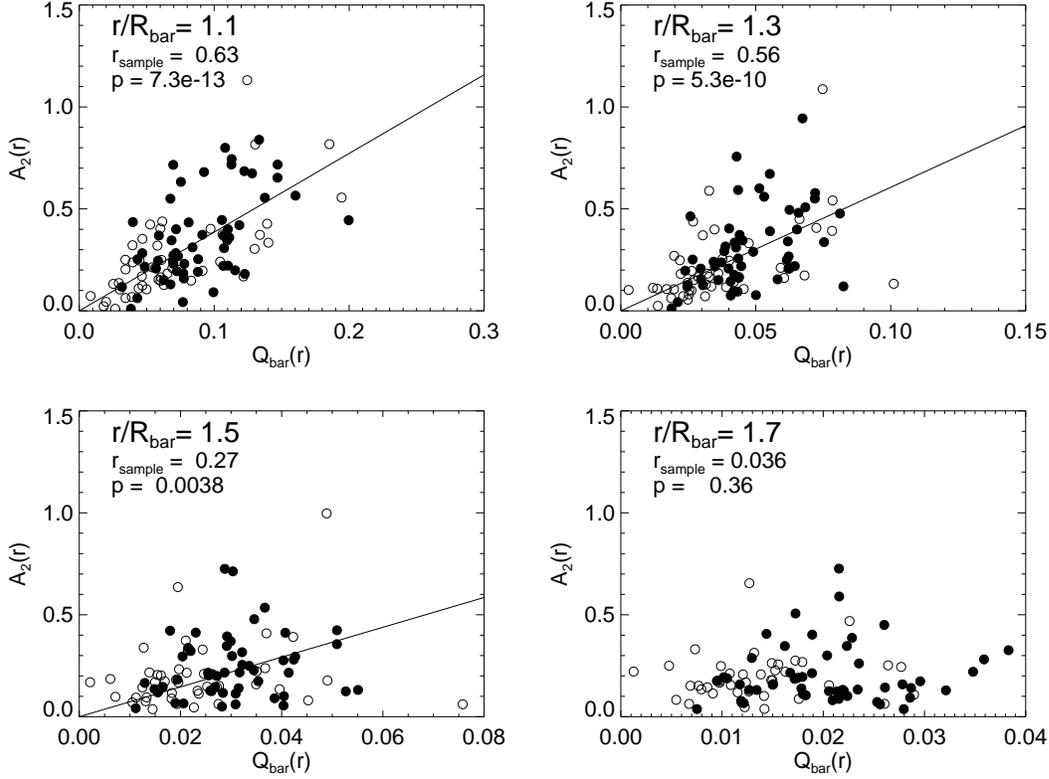}
\caption{ Relation between local bar forcing and local spiral
  amplitude.  Four distances, normalized to the bar length
  ($r/R_{bar}=1.1-1.7$) are compared, for the 103 barred
  OSUBSGS galaxies. In the calculation of bar forcing, the $m>0$
  density Fourier amplitudes are set to zero beyond $R_{cut}=
  R_{bar}$. The $p$ values indicate the significance of
  the Spearman rank correlation coefficient $r_{sample}$ (the
  probability of having $r_s> r_{sample}$ is $p$, under the
  hypothesis that the variables are independent). In the case
  where $p<0.01$, the best-fit linear relation is also indicated.  Open and
  filled circles denote short ($R_{bar}/h_R<1$) and long bars
  ($R_{bar}/h_R>1$), showing no difference.
\label{fig2}
}
\end{figure}

\clearpage

\begin{figure} 
\includegraphics[angle=0,scale=.90]{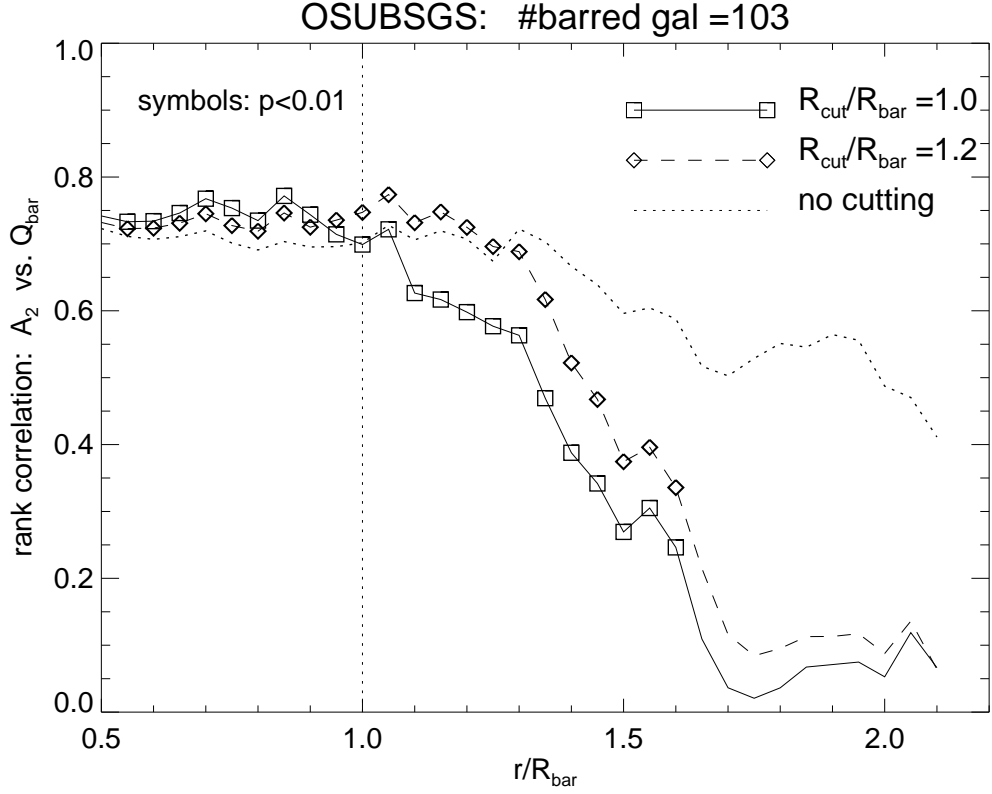}
\caption{Rank correlation coefficient between the local bar forcing
  $Q_{bar}(r)$ and the local amplitude $A_2(r)$, as a function of
  distance $r/R_{bar}$. Note that at $r/R_{bar}<1$ the parameter $A_2$
  describes the bar density contrast whereas outside the bar it
  describes the spiral arms.  Symbols indicate statistically
  significant correlation, obtained for $r \le 1.6R_{bar}$.  Note that the
  result is independent of the exact manner in which the spiral
  contribution has been eliminated from the bar forcing: compare the
  solid ($R_{cut}=R_{bar}$) and dashed ($R_{cut}=1.2R_{bar}$) curves. 
  Also shown is the case with
  no cutting (dotted line): this is just for comparison, since the
  calculated forcing is then strongly affected by the spirals
  themselves.
\label{fig3}
}
\end{figure}

\clearpage
\begin{figure} 
\includegraphics[angle=0,scale=.90]{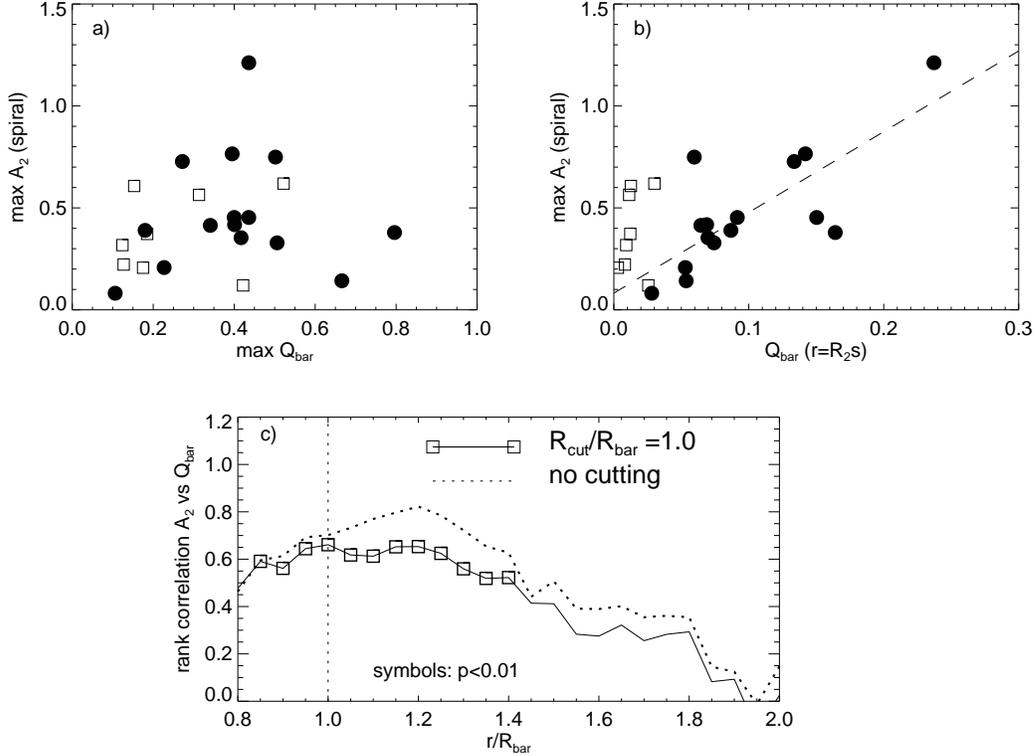}
\caption{Re-analysis of the AAT sample of 23 barred galaxies.  The
  separation of bar and spiral components in \cite{buta2009} yielded
  the maximum of bar forcing $Q_{bar}$ and the maximum of spiral
  amplitude $A_{2}$, the latter attained at distance $R_{2s}$. Frame
  {\bf a)} indicates no correlation between these quantities (same
  data as in Fig. 26a in Buta et al.). The filled and open symbols
  distinguish the galaxies with $R_{2s}/R_{bar}<1.5$ and
  $R_{2s}/R_{bar}>1.5$, respectively. {\bf b)} However, a correlation
  is present (with a significance $p=0.007$) if we compare the {\em
    bar forcing at the location} of the maximum spiral amplitude; this
  is particularly clear for galaxies with $R_{2s}/R_{bar}<1.5$ (filled
  symbols). {\bf c)} The connection of bars and spirals is even
  clearer if the local bar forcing and local spiral
  amplitude are compared as a function of distance: the figure shows the rank
  correlation coefficient vs. distance, constructed in a same manner
  as in Figs. 2 and 3.
\label{fig4}}
\end{figure}

\end{document}